\documentclass[%
 reprint,
 amsmath,amssymb,
 aps,
]{revtex4-1}

\usepackage{graphicx}
\usepackage{dcolumn}
\usepackage{bm}
\usepackage[normalem]{ulem}

\usepackage{color,ulem,soul}

\begin{document}

\preprint{APS/123-QED}

\title{The Efficacy of the Method of Four Coefficients to Determine \\ Charge Carrier Scattering}
\thanks{Electronic Supplementary Information (ESI) available}%

\author{Caitlin M. Crawford$^{1}$}
\email[Correspondence email address: ]{caitlincrawford@mines.edu}
\author{Erik A. Bensen$^{1}$}
\author{Haley A. Vinton$^{2}$}
\author{Eric S. Toberer$^{1,3}$}
\affiliation{$^1$Physics Department, Colorado School of Mines, Golden, Colorado 80401, USA.}
\affiliation{$^2$Mathematics Department, Colorado School of Mines, Golden, Colorado 80401, USA.
}%
\affiliation{$^3$National Renewable Energy Laboratory, Golden, Colorado 80401, USA.}%

\date{\today}

\begin{abstract}
The investigation of the electronic properties of semiconductors is inherently challenging due to the ensemble averaging of fundamentals to transport measurements (i.e., resistivity, Hall, and Seebeck coefficient measurements). 
Here, we investigate the incorporation of a fourth measurement of electronic transport, the Nernst coefficient, into the analysis, termed the method of four-coefficients. 
This approach yields the Fermi level, effective mass, scattering exponent, and relaxation time. 
We begin with a review of the underlying mathematics and investigate the mapping between the four-dimensional material property and transport coefficient spaces. 
We then investigate how the traditional single parabolic band method yields a single, potentially incorrect point on the solution sub-space. 
This uncertainty can be resolved through Nernst coefficient measurements and we map the span of the ensuing sub-space. 
We conclude with an investigation of how sensitive the analysis of transport coefficients is to experimental error for different sample types.
\end{abstract}

\maketitle



Holistic design strategies for novel semiconductors require an understanding on how charge carrier scattering is driven by chemical composition and microstructure.\cite{Dresselhaus2007, Snyder2008}
The challenge to date is the difficulty in resolving scattering sources and strengths as a function of charge carrier energy. Historically, most analyses prescribe a value to the energy dependence of scattering and simply incorporate scattering into a mobility value.\cite{Lundstrom2000, Wood2018} Galvanothermomagnetic measurements, such as Nernst, can shed light on scattering if combined with other thermoelectric characterization techniques.\cite{Behnia2016, Zukotynski1967} 
However, this approach is nontrivial and few groups have a history of such multi-parameter analysis.\cite{Young2003,Heremans2004,Nemov2009,Heremans2012,Demars1974} 
In this work, we provide a comprehensive study of this transformation between experimental measurables and underlying material properties.
We focus on case examples from thermoelectric materials but the results are generalizable to other classes of semiconductors.  

\begin{figure}[t]
\centering
 \includegraphics[height=4.25cm]{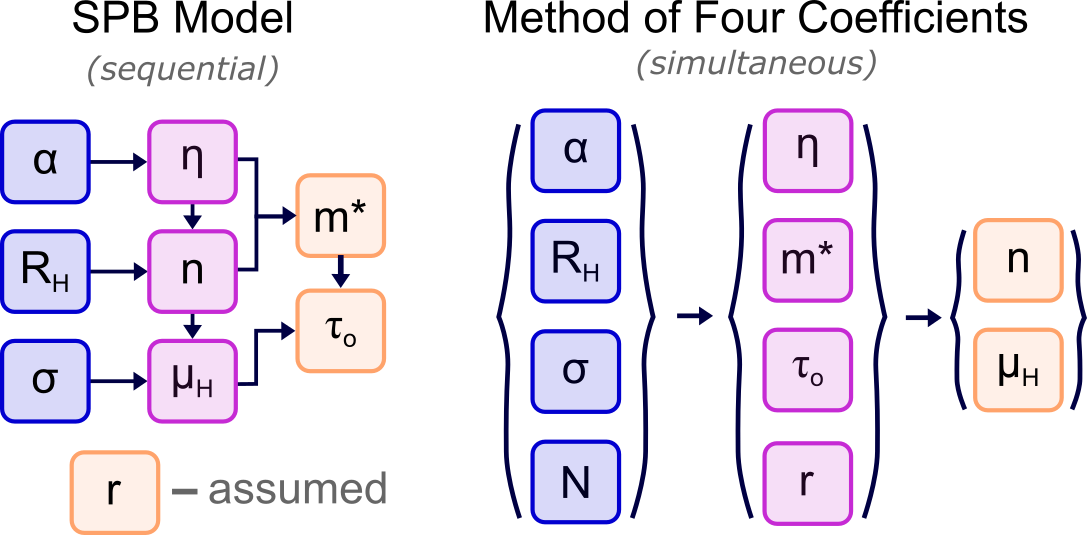}
  \caption{The single parabolic band (SPB) model begins by assuming a scattering exponent ($r$) that correlates to a dominate scattering mechanism. Sequential analysis of transport coefficients then yields the material parameters of interest. In order to eliminate this assumption and thus determine more accurate underlying material properties, a fourth coefficient (Nernst, N) can be included. As such, the method of four coefficients can be used to accurately probe scattering sources in semiconductors.}
  \label{fgr:schematic}
\end{figure}

Majority charge carrier transport is often characterized via three effects: electrical conductivity, Hall effect, and the Seebeck effect. Together these three coefficients are frequently used in conjunction to make inferences about the underlying transport phenomena.\cite{Rowe2012} 
In the field of thermoelectrics, it is common practice to assume an electronic band structure (e.g. parabolic or Kane band) and scattering type to approximate an effective mass ($m^*$), reduced Fermi level ($\eta = E_F/k_B T$ where $E_F$ is the Fermi level), and Hall mobility ($\mu_H$) for a system. The combination of these three coefficients is colloquially known as the SPB (single-parabolic band) model as seen in Fig.~\ref{fgr:schematic}.\cite{Goldsmid2010,May2009spb,Kim2015}

\begin{figure*}[t]
  \includegraphics[width=\textwidth,height=4.5cm]{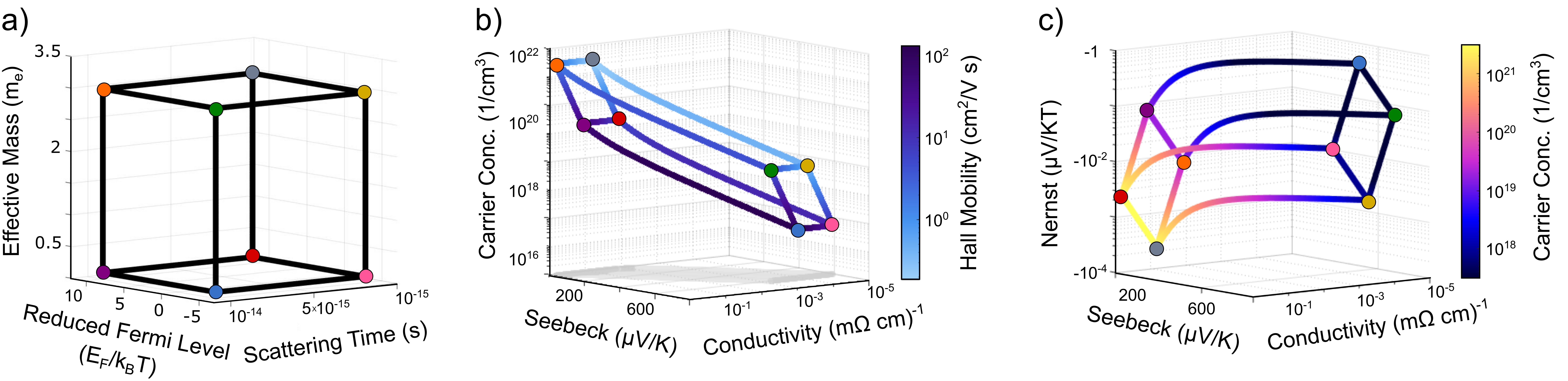}
  \caption{(a) By defining a $\mathbb{R}_3$ domain of material properties we transform this space using the  single-parabolic band model (b) to the analogous $\mathbb{R}_3$ space of transport coefficients assuming acoustic phonon scattering with $r = 0$. (c) With this same assumption, one can likewise use the method of four coefficients to transform these material properties to $\mathbb{R}_4$. Due to the difficulty of plotting a 4-dimensional space, the method of four coefficient volume has been represented here in three-dimensions with a fourth-dimension heat mapped to its transformation. For both volumes, the colored points serve as guides-to-the-eye for orienting how the vectors in the domain transform through each analysis.}
  \label{fgr:modelspaces}
\end{figure*}
 
This three-coefficient approach is favored for its simplicity and accessibility through common thermoelectric measurement techniques.  
This approach is often appropriate for parent compounds with few defects or alloying and remains a valuable method.
Further, the common use of this analysis technique has facilitated comparisons between research groups.\cite{Snyder2008}
 
Despite these attractive features, the SPB model has significant pitfalls when analyzing chemical trends.  
Specifically, changes in carrier transport are often attributed to changes in effective mass or mobility.
However, a change in the dominate scattering mechanism of the system could result in similar alterations in materials properties.\cite{Heremans2008}
This is particularly pernicious in aliovalent alloys and deleterious dopants.
As such, an approach to investigate the dominate scattering mechanism is needed.

In order to eliminate the assumption of a particular dominate scattering mechanism, a fourth measurable must be introduced to solve for our four unknown material parameters: reduced Fermi level, scattering time, effective mass, and scattering exponent ($\eta, \tau_0, m^*, r$, respectively). 
The method of four coefficients introduces a fourth transport coefficient to eliminate the assumption of scattering.
This approach manipulates Seebeck, electrical conductivity, Hall, and the transverse Nernst Effect as seen in Fig.~\ref{fgr:schematic}.\cite{Zhitinskaya1966}
An uncommon thermoelectric measurement, the Nernst effect produces a measurable transverse voltage when a longitudinal orthogonal temperature gradient and out-of-plane magnetic field are applied.\cite{Rowe2006, Tsidilkovskii}
The sign of the Nernst coefficient depends on the scattering parameter of the system and the band structure.\cite{Putley1960}
By adding this fourth coefficient to the analysis, the assumption of scattering type can be eliminated and therefore investigated. 
As such, the method of four coefficients allows us to detangle reduced Fermi level ($\eta$), density of states effective mass ($m^*$), constant relaxation time ($\tau_0$), and scattering exponent ($r$) from four experimentally measured values.   

In 1966, Zhitinskaya \textit{et. al.} introduced the concept of \textit{the method of four coefficients} to investigate the scattering effects on the non-parabolicity of the the band structure in PbTe.\cite{Zhitinskaya1966}
However, this approach has been historically under-utilized. Before 2000, most research was focused on the measurement techniques and the theory behind the method.\cite{Hall1925, Sondheimer1948, Harman1964, Putley1955, Mansfield1958, Goldsmid1972a, Demars1974, Haftman1962} Starting in the 2000s, several research groups began utilizing the method of four coefficients to perform analysis on experimental data.\cite{Jaworski2012, Young2000}

For example, in 2008, V. Jovovic used the method of four coefficients to investigate Fermi Level pinning at low-temperature in indium doped Pb-Sn-Te alloys. Ultimately, it was concluded that changes in transport were not linked to indium doping as evidenced by the lack of deviation expected in effective mass, which was analyzed via the method of four coefficients.\cite{Jovovic2008} Similar strategies to understand trends in chemical modification via this method have been used to investigate the presence of resonant energy states as evidenced by abnormal changes in effective mass.\cite{Heremans2008, Heremans2005, Heremans2012}   


Outside of thermoelectric materials, work has been done that applied the method of four coefficients to thin film transparent conducting oxide materials for photovoltaic application.
These works focused on determining which underlying material parameter limits mobility which ultimately effect optical and electrical properties.\cite{Young2000, Coutts2000, Young2000a, Young2003}

In this paper, we focus on the method of four coefficients and its effects on thermoelectric measurements.
Using a single parabolic band (SPB) assumption, we perform analysis on the 4-dimensional space and determine regions of potentially large Nernst effect signal.
In addition, we address how uncertainty in experimental measurements will affect the results of this approach.


\section{Traditional Single Parabolic Band Analysis}
The thermoelectrics community has historically focused on a 3-coefficient isotropic single parabolic band analysis, commonly referred to as the SPB model.

Here experimental measurements of Seebeck, Hall, and conductivity are used to determine Fermi Level ($E_F$), density of states effective mass ($m^*$), and Hall mobility ($\mu_H$).\cite{Rowe2012}
In order to decouple these material parameters from the measured coefficients, it is common to make various assumptions to simplify the analytics. 
It is assumed that power-law scattering is applicable and takes the form of: 

\begin{equation}
    \tau = \tau_0 \chi^{r-1/2}
    \label{eq:powerlaw}
\end{equation}

where $\tau_0$ is the constant relaxation time, and $r$ is the energy-dependant exponent which is correlated to a dominate scattering mechanism of the system. \cite{Lundstrom2000}
Here $\chi$ is the reduced energy, $\chi=E/{k_{B}T}$, where $k_B$ is Boltzmann's constant and $T$ is temperature. 
The scattering exponent (or parameter) $r$, as it is commonly called, has been calculated via Fermi's Golden Rule for common scattering types. 
Here $r = 0$ for acoustic phonon scattering, $r = 2$ for ionized impurity scattering, and $r = 1$ for polar optical phonons at high temperatures.\cite{Askerov}
 

\subsection{Governing Equations}
We can construct equations for the three coefficients (Hall, conductivity, and Seebeck) in terms of the material parameters of interest ($\eta$, $m^*$, $\tau_0$, and $r$) assuming an infinitely large band gap and one carrier type.\cite{Askerov} Gaussian units are used in this section and conversion factors to more traditional units may be found in the supplemental (ESI: Table 4). With the approximation of a parabolic band and power-law scattering (Eq. \ref{eq:powerlaw}), the electrical conductivity ($\sigma$) can be expressed in terms of an energy-dependent integral:
%
\begin{equation}
\begin{aligned}
\sigma = ne\mu = & ne^2 \left< \frac{\tau(E)}{m(E)}\right> \\
\left<A\right> =& \frac{1}{3\pi^2 n} \int_0^{\infty} \left(-\frac{\partial f_0}{\partial E} \right)k^3 (E)A(E)dE
\label{eq:condexpanded}
\end{aligned}
\end{equation}
%

where $\frac{\partial f_0}{\partial E}$ is the derivative of the Fermi-Dirac distribution with respect to energy and $k$ is the dispersion relationship of a single parabolic band ($k(E) = \left(\frac{2 m E}{\hslash^2}\right)^{1/2}$). The sign of the electric charge (e) is dictated by the carrier type ($e = -e$ for electrons and $e = +e$ for holes).
Also, $A$ is a place-holder variable representing a general function and the $< >$'s are a short hand for the integral above. 
By substitution of the power-law scattering equation (Eq. \ref{eq:powerlaw}), $\tau_0$ is pulled out of the integral and we are left with the energy dependent term.
Likewise, since we assume a single parabolic band, the effective mass is not dependent on energy allowing us to pull $(1/m^*)$ out front. Together, this gives us the following equation: 

\begin{equation}
\begin{aligned}
\sigma=& ne^2 \frac{\tau_0}{m}\left<\chi^{r-1/2}\right> \\
=&  \frac{e^2}{3\pi^2 } \frac{\tau_0}{m} \int_0^{\infty} \left(-\frac{\partial f_0}{\partial \chi} \right)k^3 \chi^{r-1/2} d\chi
\end{aligned}
\label{eq:cond}
\end{equation}

The integral of the energy dependent terms can be rewritten as a one parameter Fermi integral, $F_r$, of the form:
%
\begin{equation}
 F_r(\eta) = \int_0^{\infty}  \left(\frac{-\partial f}{\partial \chi}\right) \chi^r d\chi
 \label{eq:fermiintegral}
\end{equation}

%
where $\chi$ is the reduced energy: $\chi= \frac{E}{k_B T}$.
Extracting out the density of states from the integral, the conductivity equation in Equation \ref{eq:cond} can be rewritten:

\begin{equation}
\sigma =  \frac{e^2 (2 k_B T)^{3/2}}{3 \pi^2 \hslash^3}\,   m^{*1/2} \, \tau_0 \, \, F_{r+1}(\eta)
\label{eq:conductivity}
\end{equation}

where $\sigma$ depends on all four material parameters ($\eta, r, m^*, \tau_0$). 
Similar expressions can be obtained for the Seebeck coefficient ($\alpha$) and Hall coefficient ($R_H$) using their energy-dependent forms (ESI: Eqns. 1-7). In the case of $\alpha$, the magnitude depends strictly on $\eta$ and $r$ while being independent of $\tau_0$ and $m^*$.  

\begin{equation}
\alpha = -\frac{k_B}{e} \left(\frac{F_{r+2}(\eta)}{F_{r+1}(\eta)}-\eta\right)
\label{eq:Seebeck}
\end{equation}

 Similarly, $R_H$ is independent of $\tau_0$: 
 
\begin{equation}
R_H = -\frac{1}{e c}\left(\frac{F_{2r+1/2}(\eta)}{(F_{r+1}(\eta))^2}\right)\left(\frac{3 \pi^2 \hslash^3}{(2m^* k_B T)^{3/2}}\right)
\label{eq:Hall}
\end{equation}

%
To fit the convention of literature, the more intuitive unit of carrier concentration is used instead of the Hall coefficient.
To convert between the two for the SPB model approach, the Hall equation is solved for in terms of carrier concentration while assuming the Hall factor is unity ($a_r = 1$): 

\begin{equation}
n = \frac{a_r}{R_H e c}; ~~ a_r =\frac{F_{3/2}(\eta) F_{2r+1/2}(\eta)}{(F_{r+1}(\eta))^2}
\label{eq:carrier}
\end{equation}

For acoustic-phonon scattering $a_r = 1.18$ and neutral impurity scattering $a_r = 1.00$. A more extensive list of Hall factors may be found in ESI: Table 6. Traditional analysis assigns a Hall factor of unity even though this 3-coefficient method prescribes acoustic-phonon scattering ($r= 0$) as the dominate scattering mechanism.\cite{Lundstrom2013}  
The standard order of operations for the SPB model can be seen above in Fig.~\ref{fgr:schematic}. 
This analysis begins by assuming the dominate scattering mechanism is acoustic phonon-scattering and as such $r = 0$ remains fixed throughout this approach. 
The first step begins by solving for $\eta$ from Seebeck via Eq. \ref{eq:Seebeck}. 
Once $\eta$ is determined, the effective mass can be solved for by using $\eta$ and the carrier concentration. 
By combining Eq. \ref{eq:Hall} and Eq. \ref{eq:carrier}, an explicit form of carrier concentration can be derived:

\begin{equation}
\begin{aligned}
n =  \frac{ (2 m k_B T)^{3/2}}{3 \pi^2 \hslash^3}\, F_{3/2}&(\eta) \xrightarrow{} \\
m&= \left(\frac{n}{F_{3/2}(\eta)} \frac{\,3\pi^2\hslash^3}{(2 k_B T)^{3/2}}\right)^{2/3}
\end{aligned}
\label{eq:mstar}
\end{equation}

Finally the last parameter, the Hall mobility, is determined via the simple form of conductivity as seen in Eq. \ref{eq:condexpanded}: $\sigma = n e \mu$. We often talk about the Hall mobility ($\mu_H$) instead of the intrinsic mobility ($\mu$) though both are functions of effective mass and $\tau_0$. 
Hall mobility is often used as it lends itself easier to Hall measurements already performed.\cite{Borup2015}
Though not often reported, one could discern $\tau_0$ via the mobility by: $\mu = \left< \frac{\tau(E)}{m}\right>$.\footnote{It's important to note that this equation is for $\mu$ and not $\mu_H$ where the two differ by the Hall factor $a_r$: $\mu_H = a_r \mu$ where $a_r$ can be found in Eq. \ref{eq:carrier} (reminder $r=0$ in the SPB model).\cite{Lundstrom2013}  This becomes irrelevant in the SPB model since it is assumed there is an $a_r$ of unity which therefore leaves you with $\mu_H = \mu$.}

\begin{equation}
\tau_0= \mu\, \frac{m^*}{e} \frac{F_{3/2}(\eta)}{F_{r+1}(\eta)} 
\label{eq:mobility}
\end{equation}

As seen by equations \ref{eq:conductivity} - \ref{eq:Hall}, these three transport measurements are linked to three underlying material properties ($\eta$, $m^*$, and $\tau_0$) (when assuming a scattering exponent). As such, one could determine expected values of transport coefficients for a wide range of material parameters. 
In Fig \ref{fgr:modelspaces}a, we construct a 3-dimensional domain based on realistic ranges for such parameters: 
$1 \times 10^{-15}\,s \leq \tau_0 \leq 1\times 10^{-14}\,s $; 
$0.1\,m_e\leq m^* \leq 3.5\,m_e$;
$-5\,(E_F/k_BT)\leq \eta \leq 10\,(E_F/k_BT)$.
This domain space spans a wide region of semiconductors as evidenced by the large range of Fermi level position well-below and well-above the band edge ($\eta =0$). Also, well performing thermoelectric materials have been known to have high mobility values thus large $\tau_0$'s and small $m^*$'s. By including low values of $\tau_0$ and high values of $m^*$, we can also discern expected transport coefficients when mobility is limited.
Again, $r=0$ is assumed here for the SPB model approach. Using the ranges determined in Fig.~\ref{fgr:modelspaces}a, we solve for the three electronic transport coefficients as seen in Fig.~\ref{fgr:modelspaces}b. Here the more widely reported mobility value is heat mapped to the volume. 

Using the domain ranges as described in Fig.~\ref{fgr:modelspaces}a, the calculated coefficients take on realistic ranges. The carrier concentration spans across roughly six orders of magnitude where higher values of carrier concentration display higher values in conductivity and lower values in Seebeck, which is to be expected (Fig.~\ref{fgr:modelspaces}b: orange point). How the material parameters manifest themselves in the trend of these coefficients can demonstrated by the mobility heat map on the coefficient volume. The high mobility region exists at the blue point in Fig.~\ref{fgr:modelspaces}b. This corresponds to the same blue point in Fig.~\ref{fgr:modelspaces}a which has the expected high $\tau_0$ and low $m^*$ needed for high mobility. If we were to move across the domain space edge from the blue point to the green point in Fig.~\ref{fgr:modelspaces}a, we'd be keeping our Fermi level and scattering time fixed but increasing the effective mass. How the coefficients trend from this vector can be seen in Fig.~\ref{fgr:modelspaces}b by moving across the same direction (blue point $\rightarrow$ green point). Unsurprisingly, the mobility decreases due to an increase in the effective mass. Other trends from the material parameters to the transport coefficient can be investigated through this lens.

As evident by the lack of deformation to this volume, this transformation has unique solutions. As such, the combination of three material parameters (blue point, Fig \ref{fgr:modelspaces}a) correlates to one point in the coefficient space and vice versa (blue point, Fig \ref{fgr:modelspaces}b). 
Figure \ref{fgr:modelspaces}a-b is a visual, mathematically accurate, representation of the SPB model using Fermi-Dirac statistics.

While the SPB model is a simple analysis approach to tackle thermoelectric transport, it does have its pitfalls. Imagine a case where the Fermi level and effective mass in the system are fixed but $\tau_0$ and the scattering exponent could change. By changing $r$ you could detect an increase in the Seebeck and conductivity. Without knowledge of scattering, this effect is often linked to changes in the effective mass\cite{Pei2012, Wei2014a} when it could be a scattering effect change. In doing so, the SPB model has the potential to pigeon-hole analyses based off its assumptions of fixed scattering. We will revisit such case examples below in the following sections.

\section{The Method of Four Coefficients}

\begin{center}
  \begin{figure}[h!]
  \centering
  \includegraphics[width=7cm]{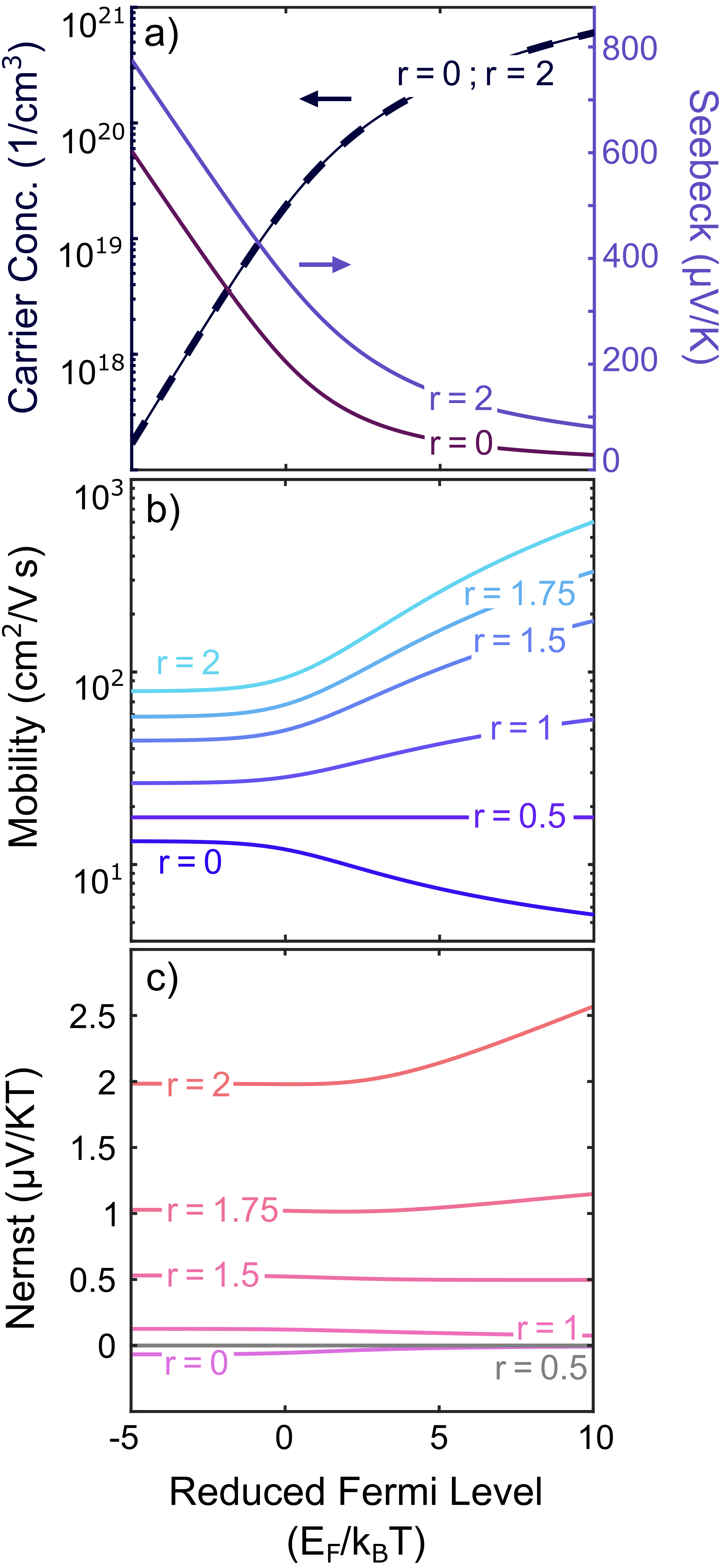}
  \caption{Trends in measurable transport coefficients are calculated as a function of reduced Fermi level ($\eta$) and various scattering exponents ($r$) (a-c). All graphs assume a single-parabolic band with transport occurring at a temperature of 300\,K and assume a scattering time ($\tau_0$ = $1\times10^{-14}$ s) and an effective mass ($m^* = 1 m_e$). As the dominate scattering mechanism changes in the system, thus changing the value of the scattering exponent, we observe expected changes in Seebeck and conductivity (a) and mobility (b). Here we see the Nernst coefficient (c), does not heavily depend on Fermi Level position but rather what scattering mechanism is dominate. Other material property trends and their effects on Nernst can be seen in the supplemental (ESI: Fig.~1c-6c).
  }
  \label{fgr:trends}
\end{figure}  
\end{center}

The SPB model measures three coefficients to solve for three unknown material parameters. 
In order to eliminate the assumption of scattering source, a fourth transport coefficient needs to be introduced to solve a system of equations with four unknowns. 
This process, known colloquially as the method of four coefficients\cite{Jaworski2012}, utilizes a different galvanothermomagentic measurement: the Nernst coefficient ($N$).
This effect, a thermal analogue to the Hall effect, generates a transverse voltage (y-direction) when an orthogonal temperature gradient (x-direction) and magnetic field (z-direction) are applied ($|N| = E_y/B_z \ T_x$).\cite{Rowe2006, Butler1940}  Though often attributed with needing high mobility for a large Nernst signal, the magnitude of the Nernst coefficient depends on all four materials parameters ($\eta$, $m^*$, $\tau_0$, and $r$): 

\begin{equation}
    N = \frac{k_B}{c}\,\frac{ \tau_0}{m^*}\,\frac{F_{r+1}(\eta)\, F_{2r+3/2}(\eta)\, - F_{2r+1/2}(\eta) \, F_{r+2}(\eta)}{(F_{r+1}(\eta))^2}
\label{eq:nernst}
\end{equation}

We see the Nernst effect scales directly with mobility prefactors ($\mu \propto \tau_0 /m^*$) but is also a function of both reduced Fermi level ($\eta$) and scattering exponent ($r$).\cite{Askerov} As such, the introduction of the Nernst effect allows us to eliminate the assumption of the scattering exponent by combining Nernst with the other three transport coefficients as seen in Fig.~\ref{fgr:schematic}b). While the SPB 3-coefficient approach had a clear order of operations to solve for the material parameters, the method of four coefficients does not. The luxury of sequential processing is lost in the method of four coefficients and all material parameters ($\eta$, $m^*$, $\tau_0$, and $r$) must be solved for simultaneously (Fig.~\ref{fgr:schematic}b). Here we provide a road map to the intricacies of the the method of four coefficients and how each transport coefficient is affected by the four underlying material parameters of interest.


\subsection{Dependencies in MO4C}

The thermoelectrics community has built an intuition of how the traditional thermoelectric effects change when we alter the Fermi level but lack intuition on how they change when scattering mechanisms.
We begin by walking through familiar trends in Seebeck and carrier concentration to orient the reader. 

In Fig.~\ref{fgr:trends}a we plot both Seebeck and carrier concentration from equations \ref{eq:Seebeck} and \ref{eq:carrier} respectively. We see that as the Fermi level rises from the band gap to deep into the band, the magnitude of the carrier concentration increases and Seebeck decreases as expected. The solid curve was generated assuming a fixed relaxation time, acoustic phonon scattering ($r=0$), an effective mass, and infinitely large band gap.  The absolute value of Seebeck is used to describe transport via electrons for clarity. Since we are no longer bound by restrictions of fixed scattering exponents, we can probe how these effects will change as a function with scattering. 

The dashed curve in Fig.~\ref{fgr:trends}a highlights how impactful different scattering effects are on the Seebeck coefficient. If we have a material system that has a Fermi Level right at the band edge ($\eta = 0$) as is typical for optimized thermoelectric materials, the difference in Seebeck between ionized impurity ($r=2$) and acoustic phonon scattering ($r=0$) differs by almost a factor of 2. Unsurprisingly, the carrier concentration is unaffected by the influence of scattering as one would expect given equation \ref{eq:carrier} does not depend on $r$. In the same vein, one could expect differences to emerge for both Hall (Eqn. \ref{eq:Hall}) and conductivity (Eqn. \ref{eq:conductivity}) when the scattering exponent changes (ESI: Fig.~4d and 4b, respectively).

In Fig.~\ref{fgr:trends}b, the log of the mobility is plotted as a function of various scattering exponents from Eqn. \ref{eq:mobility}. Again, we keep the same $\tau_0$ and $m^*$ fixed from before. Though we have fixed our mobility pre-factors ($\tau_0$ and $m^*$) it can be seen that changing scattering alone can alter mobility significantly. Here we see the highest values of mobility for ionized impurity scattering and the lowest for acoustic phonon scattering. As the Fermi level goes further into the band thus increasing carrier concentration, ionized impurity scattering increases while mobility for $r=0$ begins to decline. 

For Nernst, a consideration of Eq. \ref{eq:nernst} readily reveals the importance of large $\tau_0$ and small $m^*$ for large signal. However, the impact of $\eta$ and $r$ within the Fermi integrals are difficult to deduce by inspection.  In Fig.~\ref{fgr:trends}c,  we calculate how the Nernst coefficient changes as a function of Fermi level and various scattering mechanisms. Again, we keep $\tau_0$ and $m^*$ fixed to the same values. As one can see, for a single parabolic band the sign of Nernst only depends on the scattering exponent, $r$. For $r < 0.5$, the Nernst coefficient is negative while all values greater than 0.5 yield a positive Nernst. Material system with higher values of $r$, such is the case for ionized impurity scattering or polar optical phonon scattering, should yield the largest signals and thus make measurement viable. Nernst as a function of effective mass and scattering time may be found in supplemental (ESI: Fig.~4c-6c).

Since we assumed a scattering time and effective mass in Fig.~\ref{fgr:trends}, these are slices within the four-dimensional $\mathbb{R}_4$ range of the method of four coefficients. We will progressively expand this subspace to consider the interdependence of the transport coefficients to scattering parameter. First we consider the $\mathbb{R}_3$ (three-dimensional) space of the method of four coefficients for an assumed $r$: Fig.~\ref{fgr:modelspaces}c.

Similar to the approach taken with the SPB model in the section above, we can utilize the four equations of the electronic transport coefficients and map out the expected $\mathbb{R}_3$ space for realistic material parameters. We keep the same material parameters outlined in Fig.~\ref{fgr:modelspaces}a. For now, we hold $r$ constant at $r=0$ for easier comparison between the two models. All the colored points in Fig.~\ref{fgr:modelspaces} serve as guides-to-the-eye for the transformation process of each vector and as a result their subsequent dependencies on the transport coefficients can be examined. 

Imagine a system that exists on the blue point in Fig.~\ref{fgr:modelspaces}a. The blue point has a high $\tau_0$, low $m^*$, and a Fermi level in the band gap. We saw from equation \ref{eq:nernst} that the Nernst coefficient scales with mobility (since mobility is proportional to $\tau_0/m^*$). Indeed, we see this high mobility point in blue has a large Nernst coefficient in Fig.~\ref{fgr:modelspaces}c. Now if we were to push the reduced Fermi level into the band but keep $\tau_0$ and $m^*$ fixed, we'd be walking along the blue-to-purple point vector in Fig.~\ref{fgr:modelspaces}a. We can see how this vector is transformed by the method of four coefficients in Fig.~\ref{fgr:modelspaces}c. As we transverse the blue-to-purple vector in Fig.~\ref{fgr:modelspaces}c, we see a decrease in the magnitude of the Nernst Coefficient (for $r =0$). This is self-consistent with the trends observed in Fig.~\ref{fgr:trends}c for acoustic phonon scattering. 

\begin{figure*}
  \includegraphics[width=\textwidth]{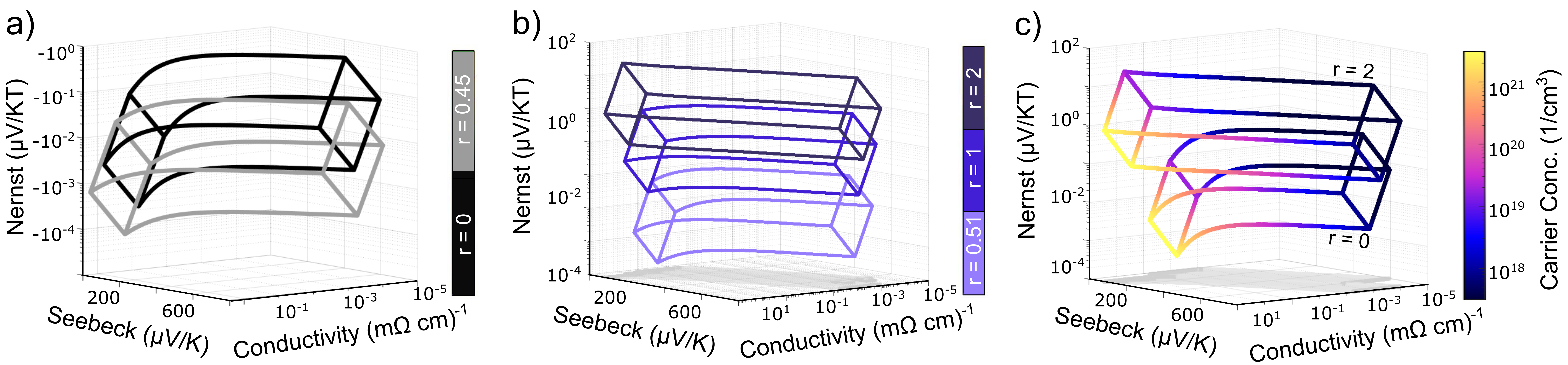}
  \caption{The four-dimensional space of the method of four coefficients has been plotted in various $\mathbb{R}_3$ sections using Fig.~\ref{fgr:modelspaces}a as the input domain and Equations \ref{eq:conductivity} - \ref{eq:nernst}. For scattering exponents less that 0.5, the Nernst coefficient will be negative for a single-parabolic band (a). All values greater than 0.5 yield a positive Nernst coefficient (b). For energy independent scattering ($r=0.5$), the Nernst coefficient goes to zero. To highlight the difference in magnitude between ionized impurity scattering ($r=2$) and acoustic phonon scattering ($r=0$) the absolute value of the Nernst coefficient has been plotted in (c) for comparison. 
  The two differ by roughly two orders of magnitude. 
  The fourth dimension, carrier concentration calculated from Hall and the correct Hall factor, has been heat mapped for clarity.
  }
  \label{fgr:4D}
\end{figure*}

In Fig.~\ref{fgr:modelspaces} we held scattering parameter constant for acoustic phonon scattering at $r = 0$. In reality, the method of four coefficients curses us with an unknown scattering mechanism. In practice $r$ can be swept between $0 < r < 2$ to see how this space evolves as a function of scattering parameter. As a result, this space is most accurately represented by a four-dimensional hypercube, a tesseract. Due to the challenges of plotting in a 4-dimensional space, instead we investigate projections in $\mathbb{R}_3$. We can plot these transformation volumes for various values of $r$ as seen in Fig.~\ref{fgr:4D}. In Figure \ref{fgr:trends}c, the sign of the Nernst coefficient switches upon crossing $r = 0.5$; Fig.~\ref{fgr:4D}a presents the case for $r < 0.5$ while  Fig.~\ref{fgr:4D}b shows the regime where $r>0.5$.
To consider the difference in magnitude between these scenarios, the two datasets are overlaid in Fig.~\ref{fgr:4D}c with the fourth dimension (carrier concentration) plotted instead. The log of Nernst has been taken to account for the sign switch between these regimes. These two scattering mechanisms are highlighted as they are often the focus of bulk thermoelectrics materials near room temperature.

Sadly, the magnitude of the Nernst coefficient will yield small measurement signals for all scattering cases as seen in Fig.~\ref{fgr:4D}. It’s particularly noxious for the case of acoustic phonon scattering since the signal is so close to zero. As such, highly accurate measurements will be needed to discern such a low Nernst coefficient. However, the differences in the sign of Nernst between acoustic phonon and ionized impurity scattering is a quick way to determine which mechanism is prominent in a system. 

By adjusting scattering parameter, we probe portions of this four-dimensional space. As can be seen in Fig.~\ref{fgr:4D}, there exists no torsion or twists to these volumes, leading us to infer this is a unique transformation. Indeed, upon inspection, no two calculated points are the same. Every point across the material property domain space we highlighted has a unique solution of transport coefficients. As such, we infer by inspection the method of four coefficients is a bijective transformation and a mathematically viable technique for determining underlying transport phenomena.

\begin{figure}[t]
  \includegraphics[width=8.75cm]{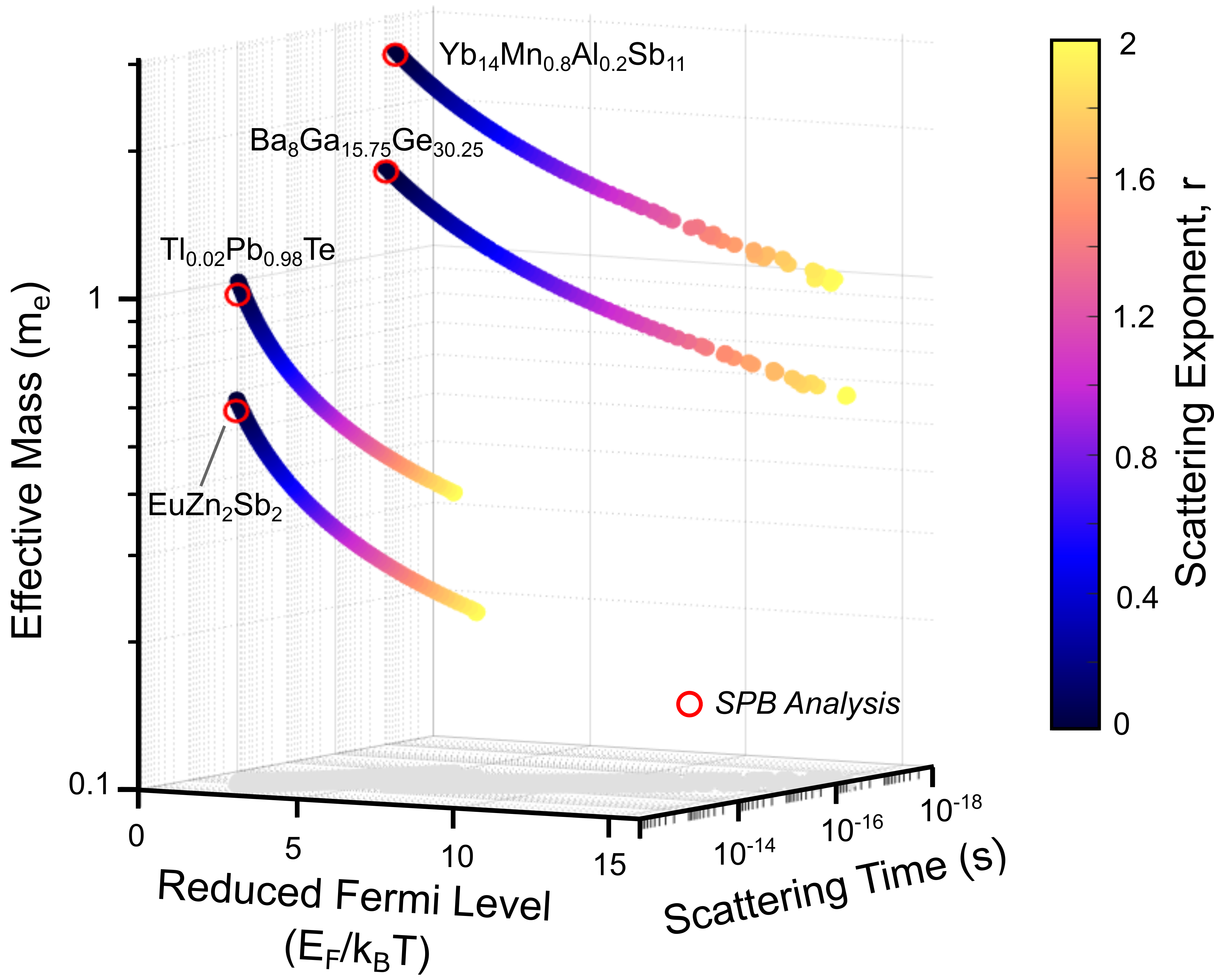}
  \caption{Here we take four material systems from literature \cite{Toberer2008spb, Toberer2010spb, May2009spb, Heremans2008} that have applied the SPB model and calculate the  method of four coefficients for all values of Nernst.
  This allows us to solve the range of material parameters that would be correct solutions for the SPB approach. Typically the SPB approach assigns $r=0$ for acoustic phonon scattering, but for these curves it spans from $r = 0$ to $2$.  The red points correspond to the properties as per calculated by the SPB model. The slight offset from the calculated data can be attributed to not using the scattering exponent dependent Hall factor (since SPB assumes $a_r=1$).}
  \label{fgr:r3r4}
\end{figure}

Having established the method of four coefficient analysis and the associated $\mathbb{R}_4 \longrightarrow \mathbb{R}_4$ transformation, we briefly investigate the uncertainty that remains in the 3-coefficient SPB analysis approach.
As a first case example, we consider our prior SPB analysis of Yb$_{14}$Mn$_{0.8}$Al$_{0.2}$Sb$_{11}$ from Ref. \citenum{Toberer2008spb}.
This sample exhibited the following properties: Seebeck = $55$\,$\mu\,V/K$ ; carrier conc. = $1.2\times10^{21}\,cm^{-3}$; resistivity = $2$\,$m\Omega\,cm$. 
By applying the SPB model to this point, these transport coefficients would yield the following material parameters as seen by the red circle labeled Yb$_{14}$Mn$_{0.8}$Al$_{0.2}$Sb$_{11}$ in Fig.~\ref{fgr:r3r4}.
However, since no information about Nernst is reported, we can hypothesize that this sample could display any reasonable Nernst values ($-0.005\,\mu\,V/K\,T$ $\leq$ N $\leq$ 0.0075\,$\mu\,V/K\,T$), moving the experimental data from a point in $\mathbb{R}_3$ to a line in $\mathbb{R}_4$.

We can then transform this $\mathbb{R}_4$ line using equations \ref{eq:conductivity}-\ref{eq:nernst} to determine how this space would be analyzed under the method of four coefficients (Fig.~\ref{fgr:r3r4}, Yb$_{14}$Mn$_{0.8}$Al$_{0.2}$Sb$_{11}$). Without information about Nernst, this $R_3$ subspace could span from the red circle all the way to the end of the line (yellow). The three traditional transport coefficients could then yield the following ranges:
$\mathbf{5.1}\leq \eta (E_F/k_BT) \leq 15.0$ ;
$1.02\leq m^* (m_e) \leq \mathbf{3.01}$ ; 
$2.7\times 10^{-17} \leq \tau_0 (s) \leq \mathbf{1.1\times 10^{-14}}$.
Here, the bold values correspond to $r=0$ while the non-bolded range is for $r=2$.  

We see the same pattern hold true for other material systems analyzed under the single parabolic band approach such as EuZn$_2$Sb$_2$, Tl$_{0.02}$Pb$_{0.98}$Te, and Ba$_8$Ga$_{15.75}$Ge$_{30.25}$.\cite{Toberer2010spb, May2009spb, Heremans2008} In all these cases, the assumption of $r=0$ results in a solution on the extreme of the possible range of material parameters.
Despite these concerns about the use of the SPB model and the associated assumption of scattering, there is a practical utility to calculating an `effective' effective mass and similar, as Nernst measurements will ultimately be quite challenging to obtain for compounds with low mobility.  However, this exercise shows that the scattering exponent used in the analysis could lead to a variation in effective mass by half an order of magnitude, a variation in reduced Fermi level by $\sim$10 (e.g. 260 meV at room temperature) and a scattering prefactor varying by three orders of magnitude.   


\section{Error Propagation of 4-Coefficient Analysis}
Though solving for a range of transport coefficients is insightful to observe trends, an experimentalist is often faced with the problem in reverse.
Sadly, the transformations from $\mathbb{R}_4$ of the transport coefficients to $\mathbb{R}_4$ of the material parameters is much more difficult. 
Instead of having the luxury of piping variables through an analytic solution, one must deconvolute the Fermi integrals and untangle the material parameters from the integrands.
To do so, numerical integration and a backsolving search process must be performed on the set of equations \ref{eq:carrier} - \ref{eq:nernst} simultaneously.
In this section we determine the material properties by their transport coefficient using the full Fermi-Dirac statistics. 
This will allow us to prescribe experimental error to the transport coefficients and probe how the error propagates through this analysis technique. 


\subsection{Methods of Numerical Integration}
It is common to use Fermi integral look-up tables to determine the value of the Fermi integrals, however this assumes one scattering exponent.
Since we are searching a continuous range of r, we calculate the Fermi integrals using the n-point Gauss-Laguerre quadrature method.
This technique allows us to numerically integrate our transport equations: Eqns. \ref{eq:conductivity}-\ref{eq:nernst}. 
This process evaluates an integral by calculating the sum of the product of the weights ($w_i$) multiplied by the integrand's $f(x)$ at a specific node ($x_i$):

\begin{equation}
    \int_0^{\infty} f(x) e^{-x} dx = \sum_{i=1}^{n} w_i f(x_i)
\label{eq:GL_sum}
\end{equation}

The nodes $x_i$ are the roots of the nth order Laguerre polynomial and the weights are computed by solving the following for each integer, k, from 0 to n-1:

\begin{equation}
    \int_0^{\infty} x^k e^{-x} dx = \sum_{i=1}^{n} w_i f(x_i)
\label{eq:fx_GL}
\end{equation}

To compute the Fermi Integrals, we used the following $f(x)$ to cancel out the $e^{-x}$ term built into the quadrature definition:

\begin{equation}
    f_k(x,\eta) =  \frac{e^{x-\eta}}{(1+e^{x-\eta})^2} x^k e^x
\label{eq:fermi_GL}
\end{equation}

 The Gauss-Laguerre quadrature computed integral of $\left(-\frac{\partial f_0(x,\eta =0)}{\partial x}\right)$ with $n =100$ matched the computed value of MATLAB's built in integrate command to machine precision. 
 As such, weights and nodes were calculated to $n = 100$ for all calculations in this paper. Open sourced Gauss-Laguerre quadrature code was adapted to fit the analysis \cite{GaussCode}. 
 
 To begin the backsolving process we define a range of guesses for each of the material parameters:  $r \in [0,2]$, $\eta \in [-5,10]\ (k_B T)$, $m^* \in [0.1,5]\ (m_e)$, $\tau_0 \in [10^{-14}, 10^{-15}]\ (s)$. These values were determined by forward solving equations \ref{eq:conductivity}-\ref{eq:nernst} for various transport coefficients to ensure our starting ranges encompassed our resulting domain. Each range was then divided into equally space points with a grid spacing of thirty. Combinations of this grid were then used to calculate a range of transport coefficients. A cost function (ESI: Eqn. 8) is then applied to each of the calculated values. This function compares the true transport coefficient values to each of the combinations calculated from our starting ranges. The point with minimum cost is then specified. 
      
To ensure we are not constraining our system to the initial starting ranges of the material parameters: if the point with minimum cost is on the edge of the starting hypercube, we shift our ranges with the point of minimum cost at the center of the ranges. Then the grid is recalculated as before. This process is performed until the point of minimum cost is no longer on an edge case of the hypercube. 

Once this is verified, the search window is centered around the point with minimum cost and we reduce the range sizes by 1/5 to calculate a finer grid search. This finer mesh goes through the same process as before stepping down continuously until the material parameter ranges ($\{\Delta r, \Delta \eta, \Delta m^*/m_e, \Delta \tau_0 \times 10^{15}\,s^{-1}\}$) can be determined within a tolerance of 0.001. The system then outputs a coordinate in $\mathbb{R}_4$ of material parameters ($r$, $\eta$, $m^*$, $\tau_0$) for the provided transport coefficients ($\alpha$, $\sigma$, $R_H$, $N$). All calculations were executed in MATLAB in Gaussian units and converted to SI units post integration.

Sequential integration of Eqs. \ref{eq:conductivity}-\ref{eq:nernst} is implemented as all transport equations are needed in determining convergence to a unique point in the material parameter space. To verify our results and ensure we are not in any local minima, we compare all numerically integrated points to the analytic forward-solving approach (as used in the sections above).  Indeed, all numerically integrated points are within machine precision to their analytic counterparts.
Therefore, all sets of transport coefficients can be used to solve for their unique material parameters (within our set of assumptions) through their full Fermi-Dirac equations by the method of four coefficients. This gives us the flexibility to forward- or backward- solve this set of equations.  


\subsection{Error in Transport Measurements}
Assessing the error when transforming from material parameters to transport coefficients can be addressed analytically (Supplemental).  Practically, however, the  back-solving process of transforming experimental transport coefficients to material parameters precludes a simple error analysis. In the following we will intentionally perturb the experimental transport coefficients with combinations of small errors. This effort relies on the unique transformation between these two spaces.  Here we prescribe the following error bars to each of the transport coefficients: $\pm 2\%$ Conductivity, $\pm 5\%$ Seebeck, $\pm 5\%$ Hall, $\pm 10\%$ Nernst plus an additional $\pm 0.05\,\mu\,V/KT$. A large Nernst error is applied since Nernst measurements are not commonly used, and the expected coefficient would yield low signals as evident by Fig.~\ref{fgr:4D}.


\begin{figure}[t]
  \includegraphics[width=7.5cm]{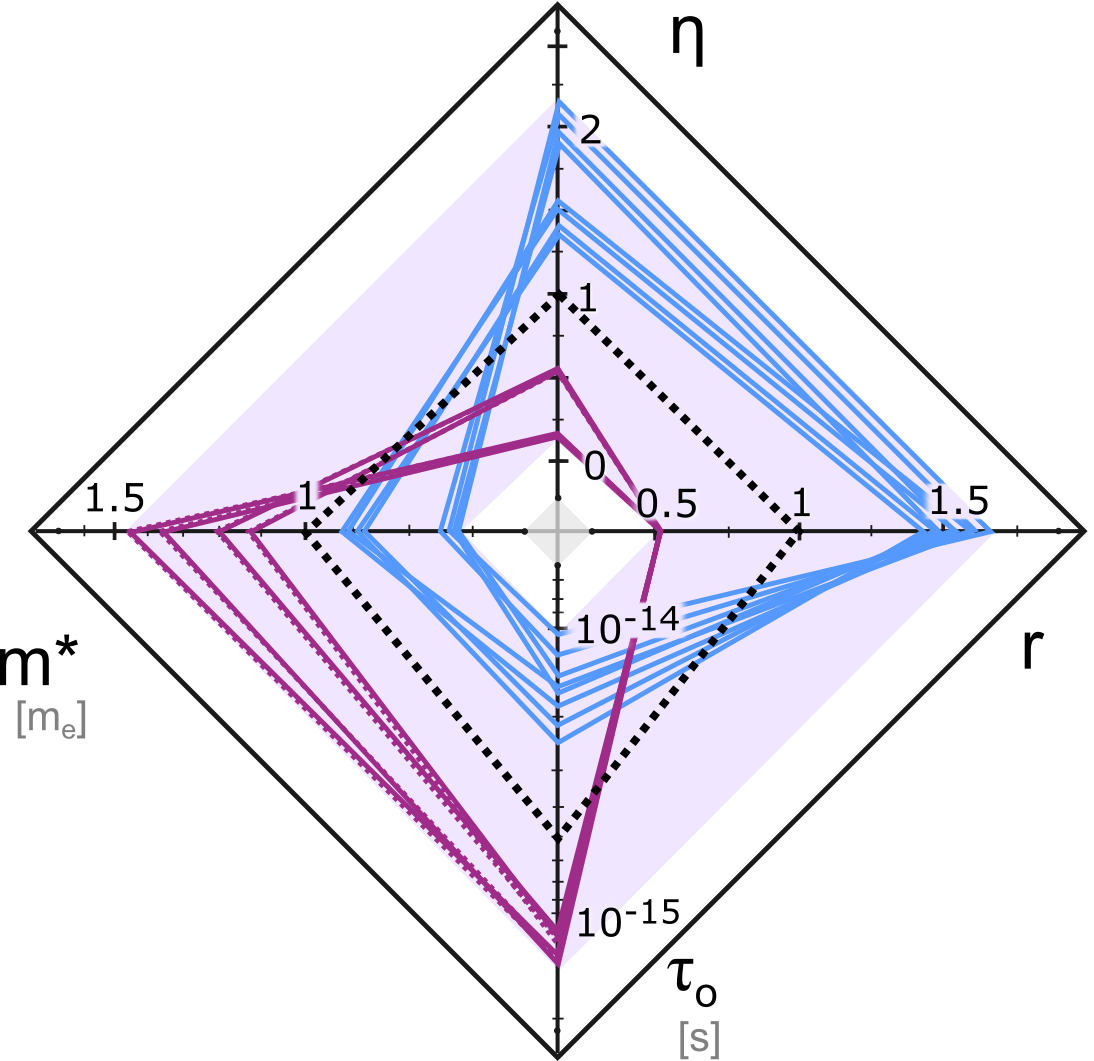}
  \caption{To investigate how experimental error propagates through the method of four coefficients, we begin with a starting seed value of material parameters (black dashed line) and analytically solve for the transport coefficients: \{$\alpha = -283\,(\mu V/K)$, $\sigma = 0.04 \,(m\Omega\,cm)^{-1}$, $N = 0.06 \,(\mu V/K\,T)$, $n$ $=1.9\times 10^{19} (1/cm^3)$\}. A systematic error is then prescribed to each of the transport coefficients and the 16 combinations of error are plotted here. This allows us to bound the region of error by the purple shaded overlay. Regions of high Nernst and low Nernst have been separated (blue and purple respectively) and its clear this causes a larger range in error as these two regions are distinct. }
  \label{fgr:spidey}
\end{figure}


\begin{figure*}[ht]
  \includegraphics[width=\textwidth]{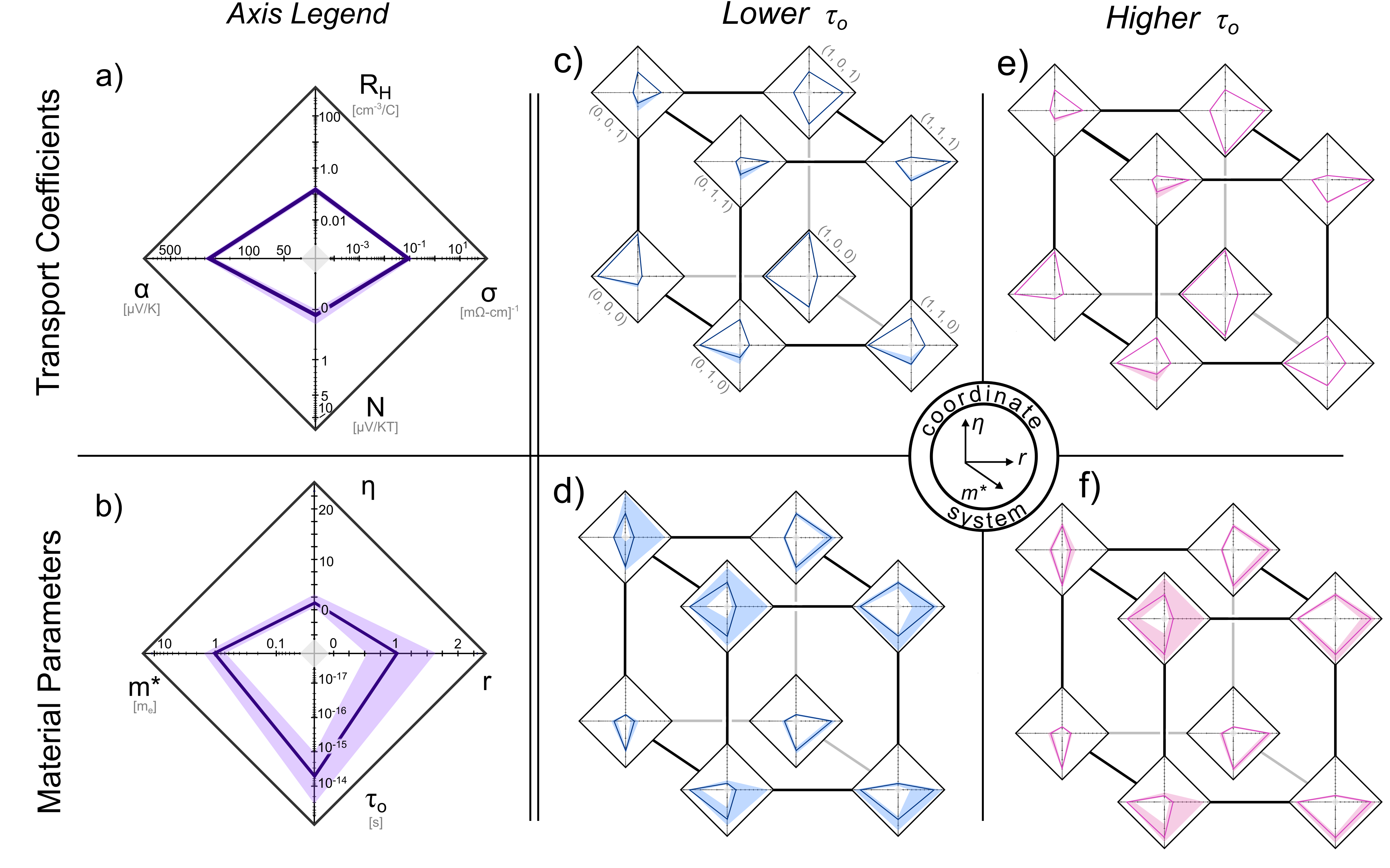}
  \caption{Experimental error is assigned to transport coefficients to observe how the method of four coefficients propagates error. Without error in the measurables (a-dark purple solid line), the method of four coefficients can solve for a unique solution for the set of scattering exponent, scattering time, effective mass, and reduced Fermi level. 
  Once experimental error is assigned to the transport coefficients (a- purple shaded region) the error in the material parameters can be numerically assessed (b- purple shaded region).  
  The purple radar plots serve as the axis legends for all the radar plots in the subsequent panels. 
  We calculate a set of experimental data points near all the extrema in Fig.~\ref{fgr:modelspaces}a and for $0 < r < 2$. 
  A standard experimental error is then applied to all coefficients and the error in material parameters are calculated. 
  The radar plots are overlaid onto their initial seed values with the coordinate system labeled. 
  Seed values with low and high $\tau_0$ have been plotted separated for clarity.}
  \label{fgr:gems}
\end{figure*}


The results of this error assessment process are shown in Fig.~\ref{fgr:spidey}.  To begin, we chose a point in our domain as a starting seed value (\{$r= 1$, $\tau_0 = 5\times 10^{-15}\,(s)$, $m^*=1\times m_e$, $\eta = 1$\}). 
This point is then used to calculate the expected experimental transport coefficients via the analytic Eqs. \ref{eq:conductivity}-\ref{eq:nernst}. In the case example considered here, this yields \{$\alpha = -283\,(\mu V/K)$, $\sigma = 0.04 \,(m\Omega\,cm)^{-1}$, $N = 0.06 \,(\mu V/K\,T)$, n $=1.9\times 10^{19} (e^-s)$\}.  The prescribed errors are then assigned to the calculated coefficients. Since there are four transport coefficients, each with their own minimum and maximum error, we calculate each combination of error leading to a set of sixteen points $(2^4)$ in this four-dimensional space. These sixteen points coarsely describe the region of potential error surrounding a set of experimental coefficients. All sixteen points and the starting coefficient value are then piped through the numerical integration process to calculate the expected value of the material properties. This can be seen by the purple shaded region in Fig.~\ref{fgr:spidey}.
 The individual combinations have been separated by high and low Nernst error by the blue and purple lines, respectively. For this case example, the assigned error in coefficients yield the following ranges in material properties: {$(0.51 < r < 1.67)$, $(0.17 < \eta < 2.19)$, $(0.72 < m^* < 2.19)$, $(1.03\times10^{-15} < \tau_0 < 1.33\times10^{-14})$}. 

The $\mathbb{R}_4$ tesseract of material properties was found to have widely varying error propagation when applying the method of four coefficients. To explore this range, Fig.~\ref{fgr:gems} traverses this tesseract using the same notation as found in Fig.~\ref{fgr:spidey} (these same results are replicated in Fig.~\ref{fgr:gems}a-b as a legend).  Note, the axes have been expanded in Fig.~\ref{fgr:gems}b and the figure has been simplified to show only the ``true'' value and the range of error, denoted by the purple shaded region.  Note the Nernst axis is on a cube root as it spans both positive and negative values.
The solid purple lines of Fig.~\ref{fgr:gems}a-b are the direct transform of the other. 

Fig.~\ref{fgr:gems}c-f shows a selection of points across the tesseract to see trends in the error. Here we have chosen seed values near all the extrema of $\mathbb{R}_3$ in Fig.~\ref{fgr:modelspaces}a and $r=0,2$.  The two rows describe the transport coefficients and material parameters, respectively, while the two columns involve systems with low and high $\tau_0$.  Within each panel, the cube is decorated with eight diamonds, the legends for which are shown in panel (a) and (b).  The coordinate system within each cube is shown by the axes in the center. In the following we will use a simple indexing (shown explicitly in Fig.~\ref{fgr:gems}c) where the origin is (0,0,0) and corresponds to low effective mass, small $r$ and low reduced Fermi level.  

Fig.~\ref{fgr:gems} reveals regions of largest error are associated with low $\tau_0$ and large $m^*$ (e.g. corner (0,1,1) of panel d).  Conversely, the error decreases for large $\tau_0$ and low $m^*$ (e.g. corner (0,0,1) of panel f). This trend is primarily due to the dependence of the Nernst coefficient magnitude on these two parameters (Eq. \ref{eq:nernst}). Similar trends can be seen for the pair of panels c-(0,1,0) with  f-(0,0,0) and related pairings.  

The reduced Fermi level and scattering exponent also play significant roles in determining the error in the method of four coefficients.  As seen in Figure \ref{fgr:trends}c, for $r<1$, the magnitude of the Nernst coefficient is quite small. The impact of such a weak Nernst signal can be generally seen by direct comparison of the left and right sets of points, i.e. $(h,k,0)$ vs $(h,k,1)$.  In all cases, the increase in $r$ reduces the error. This issue is further compounded at high $\eta$ for samples with small $r$.  Figure \ref{fgr:trends}c highlights that these compounds exhibit exceptionally weak Nernst signals, leading to significant error in data analysis. For example, point $(0,0,1)$ in panel (d) has significant uncertainty due to the small value of $N$ at large $\eta$ for $r=0$.  Transitioning to $r=2$ eliminates this null value for $N$ and dramatically decreases the error.  For comparison, change in $r$ created by transitioning from panel (d) (0,0,0) to (1,0,0) shows minimal error in both samples due to the strong $N$ signal for $r=2$.
 

\section*{Conclusions}
Understanding charge carrier dynamics in semiconductors is inherently difficult due to the ensemble averaging in common measurements. Further compounding the problem is that semiconductors possess a rich array of scattering sources (various defects, spectrum of phonons, etc.) that contribute to charge carrier scattering. As such, there is a persistent need to more accurately model the energy dependence of charge carrier scattering if we are to understand the connection between chemical structure and transport physics. Nernst coefficient measurements have sporadically been conducted over the last century and half and have occasionally been incorporated into multi-component analytic approach known as the method of four coefficients. In this work, we explore the inter-dependencies between the four material parameters and their four associated transport coefficients.
Within the parabolic band assumption, the $\mathbb{R}_4$ to $\mathbb{R}_4$ transformation is found to be mathematically unique. 
Our exploration of the transformation space provides guidance to experimentalists on the expected magnitudes of the pertinent transport coefficients as a function of material properties. 
Case examples of materials previously analyzed by the single parabolic band model are re-analyzed to determine the possible solution space that could have been obtained with Nernst measurements. We find the assumption of scattering coefficient traps analysis to a small subset of the potentially correct solution space. 
Having shown the potential for the method of four coefficients to successfully distinguish between changes in $m^*$, $\eta$, $\tau_0$ and $r$, we consider the impact of experimental error on such analysis. The error propagation revealed that areas with low mobility yield large error due to the low value of $N$ and that high $\eta$ and small $r$ can be challenging as well.  While the method of four coefficients doesn't overcome the challenge of ensemble averaging of charge carrier dynamics, we ultimately conclude the inclusion of Nernst measurements and the associated method of four coefficients provides critical insight into charge carrier scattering.


\begin{acknowledgments}
This work was performed at the California Institute of Technology/Jet Propulsion Laboratory under contract with the National Aeronautics and Space Administration. This work was supported by the NASA Science Mission Directorate's Radioisotope Power Systems Thermoelectric Technology Development Project under Grant/Contract/Agreement No. NNX16AT18H and No. 80NSSC19K1290. We also acknowledge support from the National Science Foundation (NSF) Grants \#1555340 and \#1950924.
\end{acknowledgments}

\bibliographystyle{rsc}
\bibliography{main}

\end{document}